\documentclass[twocolumn,prb,showpacs,multicol,amsmath,amssymb]{revtex4-1}

\usepackage{graphicx}
\usepackage{epstopdf}
\usepackage{bm}% bold math
\usepackage{subfigure}
\usepackage{sidecap}
\newcommand{\si}{\sigma}
\newcommand{\al}{\alpha}

\newcommand{\ka}{\kappa}

\newcommand{\bsig}{{\bm\sigma}}

\newcommand{\be}{\begin{equation}}
\newcommand{\ee}{\end{equation}}

\newcommand{\bea}{\begin{eqnarray}}
\newcommand{\eea}{\end{eqnarray}}
\newcommand{\bd}{\begin{displaymath}}
\newcommand{\ed}{\end{displaymath}}
\newcommand{\ba}{\begin{array}}
\newcommand{\ea}{\end{array}}
\newcommand{\bi}{\begin{itemize}}
\newcommand{\ei}{\end{itemize}}
\newcommand{\bc}{\begin{center}}
\newcommand{\ec}{\end{center}}
\newcommand{\bfl}{\begin{flushleft}}
\newcommand{\efl}{\end{flushleft}}
\newcommand{\bfr}{\begin{flushright}}
\newcommand{\efr}{\end{flushright}}
\newcommand{\non}{\nonumber}

\newcommand{\hG}{\hat{G}}

\newcommand{\hU}{\hat{U}}

\newcommand{\hbz}{\hat{{\bf z}}}

\newcommand{\tL}{\tilde{\Lambda}}

\newcommand{\om}{i\omega_n}
\newcommand{\e}{\epsilon}

\newcommand{\CC}{CeCoIn$_5$}
\newcommand{\SR}{Sr$_2$RuO$_4$}

\def\dg{^{\dagger}}

\def\bk{{\bf k}} \def\bq{{\bf q}}

 \def\bd{{\bf d}}  
  \def\hbz{\hat{{\bf z}}}

\def\da{\downarrow} \def\ua{\uparrow}
 
\def\dg{\dagger}

\def\={\!\!\!&=&\!\!\!}
\def\+{\!\!\!&&\!\!\!+~}
\def\-{\!\!\!&&\!\!\!-~}

\begin{document}
\date{\today}
\title{Multiorbital and  hybridization effects in the quasiparticle interference of triplet superconductor Sr$_2$RuO$_4$}

\author{Alireza Akbari$^1$}
\author{Peter Thalmeier$^2$}
\affiliation{$^1$Max Planck Institute for Solid State Research, D-70569 Stuttgart, Germany
\\
$^2$Max Planck Institute for the  Chemical Physics of Solids, D-01187 Dresden, Germany}

 \begin{abstract}
 The tetragonal compound \SR~ exhibits a chiral p-wave superconducting (SC) state of its three $t_{2g}$-type 
 conduction bands. The characteristics of unconventional gap structure are known from experiment, in particular 
 field-angle resolved specific heat measurements and from microscopic theories. A rotated extremal structure
 on the main active SC band with respect to the nodal gaps on the passive bands was concluded. We propose that
 this gap structure can be further specified by applying the STM- quasiparticle interference (QPI) method. We calculate
 the QPI spectrum within a three band and chiral three gap model and give closed analytical expressions. 
 We show that as function of bias voltage the chiral three gap model will lead to characteristic changes in QPI that 
 may be identified and may be used for more quantitative gap determination of the chiral gap structure.
 \end{abstract}


\maketitle

\section{Introduction}
The quasi-2D compound \SR~ is one of the few established cases of triplet superconductivity \cite{mackenzie:03,maeno:12}. This assignment follows from experimental facts like the absence of Knight shift \cite{ishida:98,ishida:01}, presence of a spontaneous condensate moment (time reversal symmetry breaking) \cite{luke:98,Gradhand:2013}, evidence for a superconducting two component order parameter \cite{kealey:00} and the absence of a Hebel-Slichter peak \cite{ishida:97}. The unconventional nature of the order parameter is also witnessed by field-angular resolved specific heat \cite{deguchi:04} and thermal conductivity \cite{izawa:01} investigations in the vortex phase. These results imply a multiband nodal triplet superconducting order parameter \cite{deguchi:04a}.

The conduction band structure is formed by the three $t_{2g}$ orbitals of $xz,yz$ and $xy$ type where the former hybridize with each other. It was found  \cite{deguchi:04,maeno:12,nomura:02,nomura:05,annett:03} that the gap structure consists of a main gap on the `active' $xy$-type band which is nodeless but has deep minima along [100] directions and smaller (nearly) nodal gaps on the `passive' $xz,yz$ bands with minima or nodes along [110] directions. 

Although the basic gap features are clear it would nevertheless be desirable to confirm and specify the gap model further. The STM technique has recently proved quite powerful in this respect for strongly correlated unconventional superconductors. The determination of the tunnelling conductance map on a finite surface area leads, via Fourier transformation, to the quasiparticle interference (QPI) pattern that is caused by impurity scattering on the surface. This pattern contains information on the normal state conduction band Fermi surface as well as on the superconducting gap structure \cite{capriotti:03, balatsky:06, maltseva:2009}. In particular for bias voltage smaller than the gap amplitude it leads to characteristic QPI features at wave vectors that allow conclusions on the \bk - dependence of the gap function. 
This technique has been used successfully to investigate the superconducting gap structure of cuprates \cite{Hoffman:2002,*Wang:2003,*mcelroy:2003,*pereg:08}, Fe-pnictides \cite{Hanaguri:2010,*Chuang:2010,*Akbari:2010,*Knolle:2010,*allan:12,*zhang:09,*huang:11} and more recently heavy fermion compounds \cite{akbari:11,allan:13,zhou:13}.\\

We believe it could also be used for the further analysis of the chiral p-wave gap structure in \SR. In fact the quasiparticle density of states (DOS) (which is the integral over the QPI function) has already been investigated in \SR~ recently \cite{firmo:13} demonstrating the feasibility of this approach. Before it was also applied to the non-superconducting Sr$_3$Ru$_2$O$_7$ compound \cite{lee:09,lee:10}

In this work we therefore investigate the expected QPI momentum and frequency structure for \SR~ in detail. We start from the known parameterization of conduction bands and use a simple representation of all three gap functions on active and passive bands that reproduce the microscopic gap structure determined by Nomura \cite{nomura:02,nomura:05} quite reasonably, in particular its extremal and nodal structure.
Using this well defined model we calculate the expected QPI spectrum as function of bias voltage. We perform a fully analytical 
calculation in Born approximation and give a closed solution for the QPI spectrum, including the subtle effects of hybridization in the passive bands. Our approach is complementary to the full t-matrix numerical treatment \cite{gao:13} which also uses different gap models.

We show that due to the 
multiband gaps and the rotated gap extrema on active and passive bands typical changes in the QPI spectrum are to be expected
when the bias voltage changes. We identify the characteristic wave vectors that appear in QPI from comparing to the structure of 
constant quasiparticle energy surfaces. These features, if compared to future experimental determination of QPI, may be used to 
further quantify the known superconducting gap structure of \SR.

\label{sec:introduction}

%%%%%%%%%%%%%%%%%%%%%% figure %%%%%%%%%%%%%%%%%%%%%%%%%%%%%%%%%%%%%%%%%%%%%
\begin{figure}[t]
\includegraphics[width=0.98\linewidth]{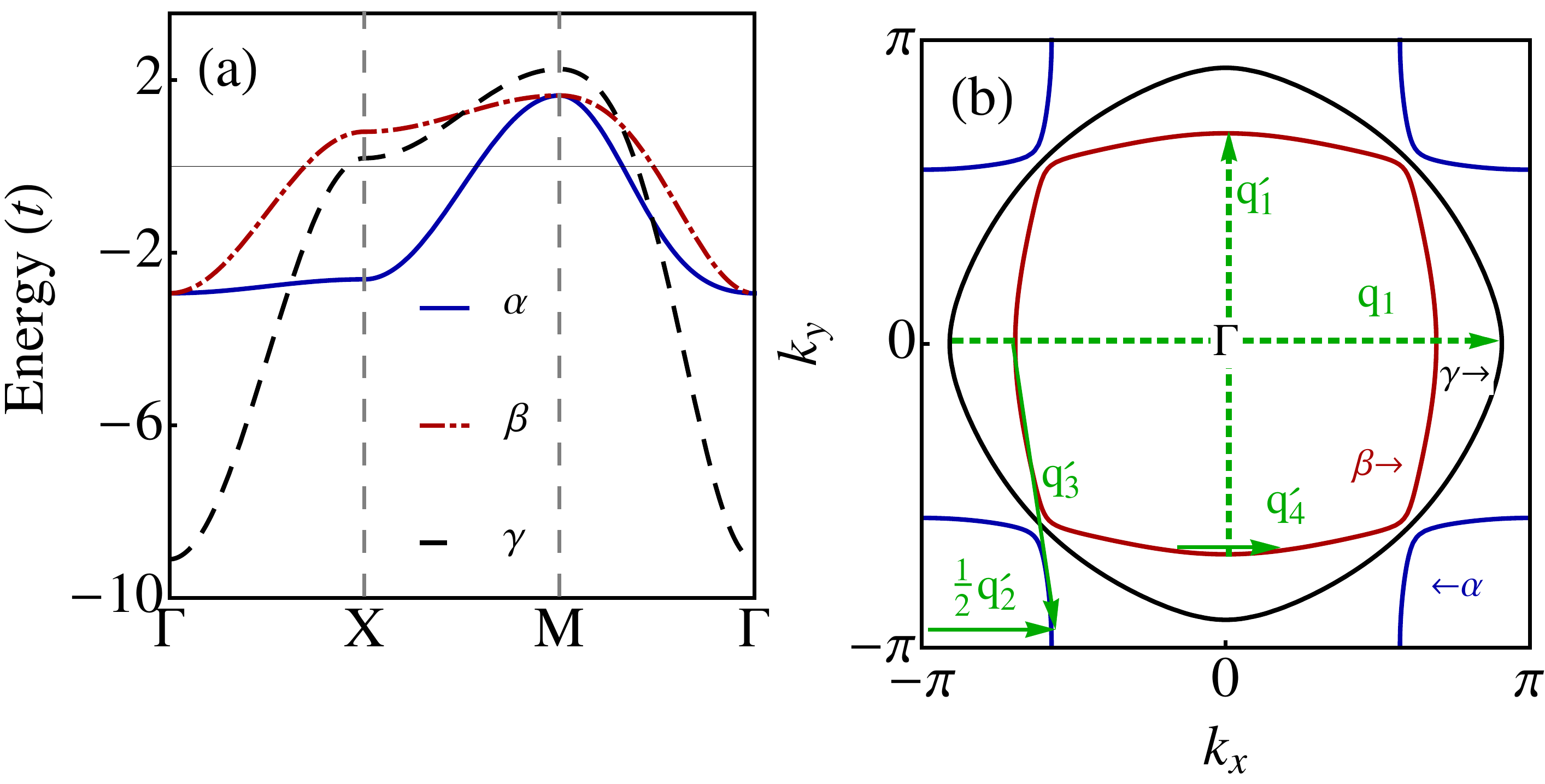}
\caption{(a) Hybridized $E_{1,2}(\bk)$ ($\alpha$,$\beta$) and unhybridized $E_3(\bk)$ ($\gamma$) band dispersions according to Eq.~(\ref{eq:hybands}).
(b) Fermi surface sheets of \SR~ with one unhybridized band ($\gamma$) resulting from xy (=c) orbitals and two hybridized  ($\alpha,\beta$) bands resulting from xz (=a) and yz (=b) orbitals. Parameters are given below Eq.~(\ref{eq:bands}).
Characteristic QPI wave vectors $\bq_i$ and $\bq'_i$ for $\gamma$ and $\alpha,\beta$ bands are indicated (c.f. Fig.~\ref{fig:Fig4}).}
\label{fig:Fig1}
\end{figure}
%%%%%%%%%%%%%%%%%%%%%%%%%%%%% figure %%%%%%%%%%%%%%%%%%%%%%%%%%%%%%%%%%%%%%%

\section{Three orbital model of quasi- 2D electronic bands in \SR}
\label{sec:bandmodel}

The quasi-2D bands of \SR~ originate from the three $t_{2g}$ 3d orbitals $d_{xz}$, $d_{yz}$ and $d_{xy}$ which are denoted by 
$n =a,b,c$ respectively. The effective tight binding (TB) model Hamiltonian for these states may be defined as \cite{liebsch:00,eremin:02,raghu:10, hughes:13}
\bea
H_0&=&\sum_{\bk,n,m}h_0^{nm}(\bk)c_{n\bk}^\dagger c_{m\bk}; 
%\;\;\;
\nonumber\\
h_0({\bf  k})&=&
\left[
 \begin{array}{ccc}
 \epsilon_{xz}(\bk) & V(\bk) & 0\\
V(\bk) &   \epsilon_{yz}(\bk)& 0\\
0 & 0 &  \epsilon_{xy}(\bk)
\end{array}
\right],
\label{eq:hmat}
\eea
where $c_{n\bk}^\dagger$ creates the unhybridized conduction electrons.
Their  dispersion $\epsilon_n(\bk)$ for each orbital and their hybridization $V(\bk)$ are parametrized as 
\bea
\e_{a\bk}=\e_{xz}(\bk)&=&-\e'_0-2t\cos k_x -2t_\perp\cos k_y, \non\\
\e_{b\bk}=\e_{yz}(\bk)&=&-\e'_0-2t\cos k_y -2t_\perp\cos k_x , \non\\
\e_{c\bk}=\e_{xy}(\bk)&=&-\e_0-2t'(\cos k_x+\cos k_y)
\nonumber\\
&&+4t''\cos k_x \cos k_y,
 \non\\
V(\bk)&=&V_\bk=-2V_m\sin k_x\sin k_y.
\label{eq:bands}
\eea
\SR~ is nearly 2D therefore dispersion along $k_z$ is neglected. The in-plane parameters are chosen as in Ref. \onlinecite{eremin:02}: $(\e'_0= 0.77, t=1.0 , t_\perp = 0.14)$,  $(\e_0= 1.61, t'= 1.39, t''= 0.45)$ and $V_m=0.1$. Absolute energy unit is $t$. Within LDA approximation it is given by $t \simeq 0.3$ eV \cite{liebsch:00, eremin:02}. The effective bandwidth or effective $t$ is however reduced by a factor of 3.5 due to correlations \cite{mravlje:11}, i.e. to $t=0.085$ eV. The spin-orbit coupling \cite{veenstra:13} is neglected in our model since we are only interested in the charge QPI without resolving the spin channels in the conductance.
The TB Hamiltonian may easily be diagonalized to give the three conduction bands
\bea
E_{1}(\bk)&=&\frac{1}{2}(\e_{a\bk}+\e_{b\bk})-
\frac{1}{2}[(\e_{a\bk}-\e_{b\bk})^2+4V_\bk^2]^\frac{1}{2}\non\\
E_{2}(\bk)&=&\frac{1}{2}(\e_{a\bk}+\e_{b\bk})+
\frac{1}{2}[(\e_{a\bk}-\e_{b\bk})^2+4V_\bk^2]^\frac{1}{2}\non\\
E_3(\bk)&=&\e_{c\bk}.
\label{eq:hybands}
\eea
Here $E_{1,2}(\bk)$ are hybridized 2D bands resulting from an anti-crossing of quasi-1D a,b bands along $(\pm\pi,\pm\pi)$ directions. Furthermore $E_{3}(\bk)$ is the unhybridized 2D xy band. The correspondence to conventional band notation is given by $(1,2,3)\equiv (\alpha,\beta,\gamma)$. Their dispersions as obtained from the model described above are shown in Fig.~\ref{fig:Fig1}a.
The three conduction bands were determined in ARPES experiments \cite{deguchi:04a,damascelli:00} and their associated Fermi surface sheets are shown in Fig.~\ref{fig:Fig1}b. The hybridized dispersions fulfil the identity
\bea
E_{1\bk}+E_{2\bk}=\e_{a\bk}+\e_{b\bk};\;\;\; E_{1\bk}E_{2\bk}=\e_{a\bk}\e_{b\bk}-V_{\bk}^2.
\eea
In the limit of vanishing hybridization $V_\bk \rightarrow 0$ the hybridized bands are given by 
$E_{1\bk}=\e_{a\bk}-(\e_{a\bk}-\e_{b\bk})\theta_H(\e_{a\bk}-\e_{b\bk})$ and
$E_{2\bk}=\e_{b\bk}+(\e_{a\bk}-\e_{b\bk})\theta_H(\e_{a\bk}-\e_{b\bk})$
where $\theta_H(\ldots)$ is the Heaviside function. Therefore a small hybridization rearranges corrugated quasi-1D Fermi surface (FS) sheets of 
$\e_{a\bk},\e_{b\bk}$ which are parallel to $k_y,k_x$ respectively into the square shaped 2D FS sheets of the hybridized $E_{1\bk}, E_{2\bk}$ ($\alpha$,$\beta$) bands shown in Fig.~\ref{fig:Fig1}b. We note that from Fig.~\ref{fig:Fig1}b the curvature of $\alpha$ or $\beta$ and $\gamma$ sheets (implying a relative rotation by $\pi/4$) are quite similar. Therefore there is no reason to make
a fundamental distinction concerning their quasi-2D character.

%%%%%%%%%%%%%%%%%%%%%% figure %%%%%%%%%%%%%%%%%%%%%%%%%%%%%%%%%%%%%%%%%%%%%
\begin{figure}[t]
\includegraphics[width=0.98\linewidth]{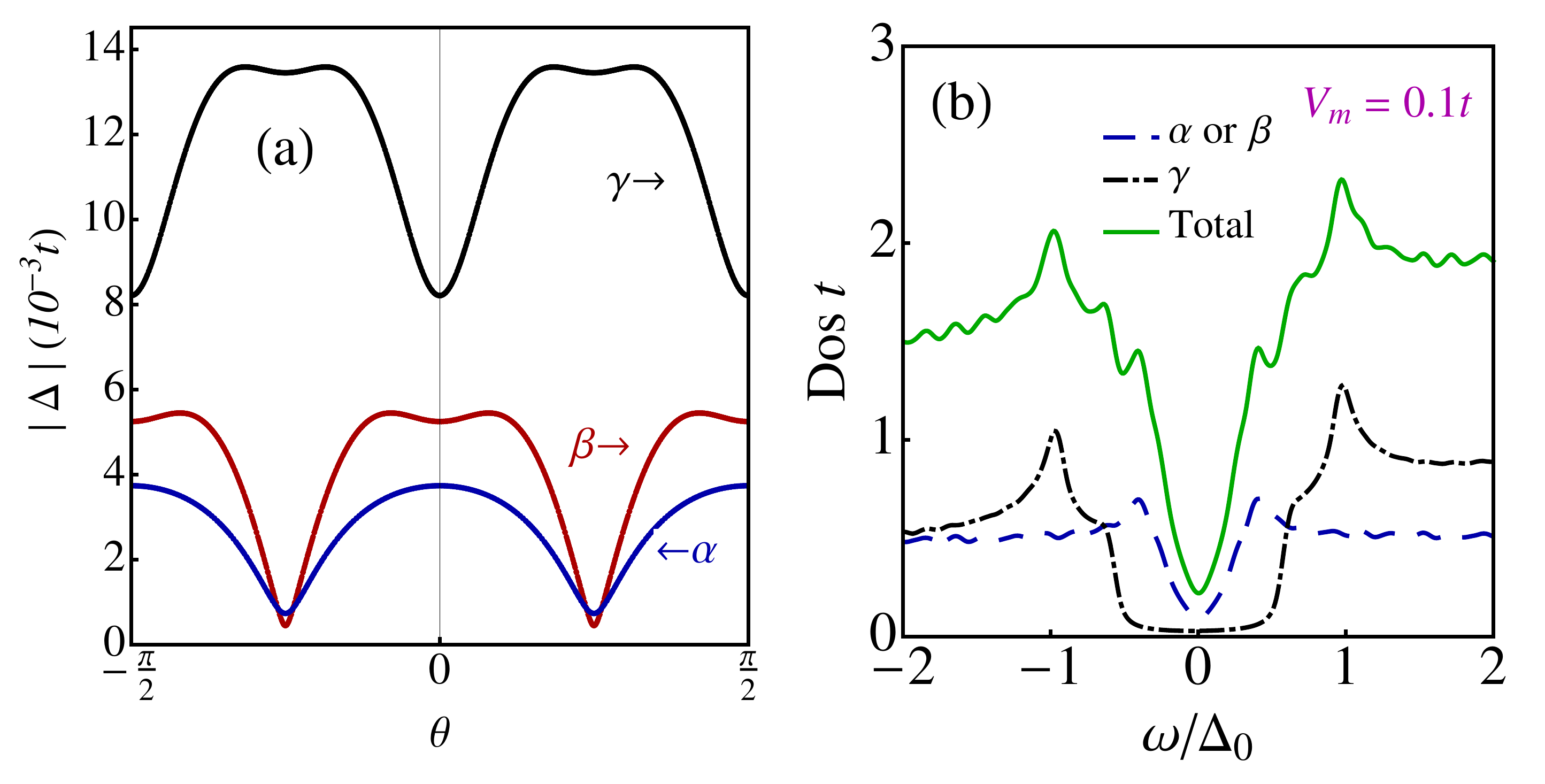}
\caption{(a) The variation of superconducting gap $\Delta_n(\bk)$ $(n=\alpha,\beta,\gamma)$ on the {\it hybridized} Fermi surfaces as function of azimuthal angle $\theta=\tan^{-1}(k_y/k_x)$ counted from the $\Gamma (0,0)$ point for $\beta,\gamma$ and from the $M (\pi,\pi)$ point for $\alpha$. Gap parameters are $\Delta_0=0.045t $, $\Delta'_0=0.01t$ and $A=0.98$, $A'=-0.7$ ($|\Delta_{\alpha_{max}}|\sim 0.004t $, $|\Delta_{\beta_{max}}|\sim0.006t $  and , $|\Delta_{\gamma_{max}}|\sim0.014t$).
(b) Quasiparticle DOS in the superconducting state. The asymmetry is due to the underlying normal state DOS.}
\label{fig:Fig2}
\end{figure}
%%%%%%%%%%%%%%%%%%%%%%%%%%%%% figure %%%%%%%%%%%%%%%%%%%%%%%%%%%%%%%%%%%%%%%

%
%%%%%%%%%%%%%%%%%%%%%% figure %%%%%%%%%%%%%%%%%%%%%%%%%%%%%%%%%%%%%%%%%%%%%
\begin{figure}[t]
\includegraphics[width=0.98\linewidth]{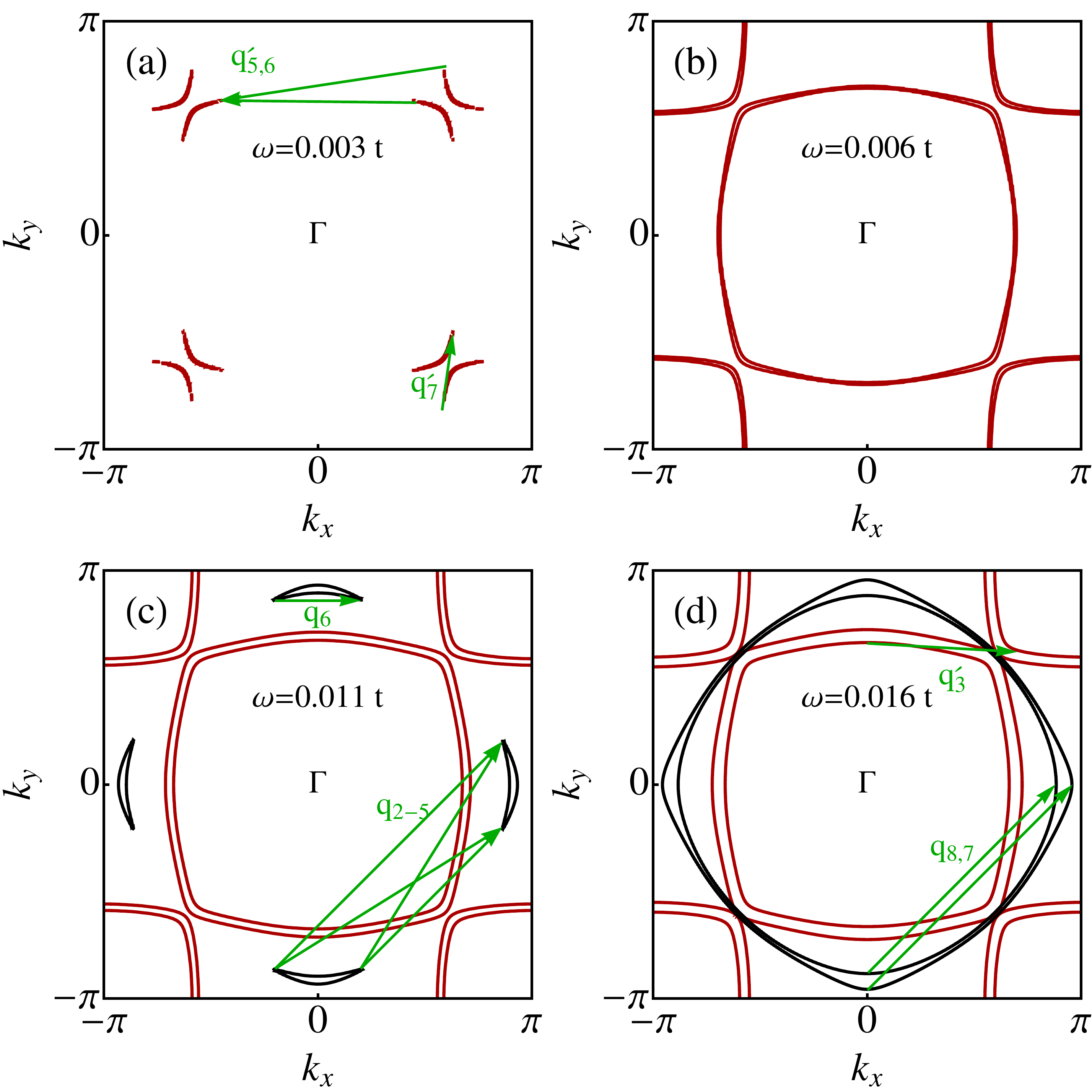}
\caption{Surfaces of constant quasiparticle energy $\Omega_{n\bk}=\omega$ (n=1-3) for various $\omega$ in the superconducting state (c.f. Fig~\ref{fig:Fig2}a). For small $\omega$ only surfaces connected with $\alpha,\beta$ bands are present, first as arcs around [110]-type directions. 
For larger energies the surfaces of $\gamma$ band also appear , first as lenses along [100]-type directions rotated by $\pi/4$ with respect to low energy $\alpha,\beta$ sheets. Characteristic QPI wave vectors $\bq_i$ and $\bq'_i$ for $\gamma$ and $\alpha,\beta$ bands are indicated (c.f. Fig.~\ref{fig:Fig5}).}
\label{fig:Fig3}
\end{figure}
%%%%%%%%%%%%%%%%%%%%%%%%%%%%% figure %%%%%%%%%%%%%%%%%%%%%%%%%%%%%%%%%%%%%%%
%

\section{The  chiral p-wave superconducting gap function of \SR}
\label{sec:gapmodel}

There are numerous SC gap models that have been discussed for \SR~\cite{rice:95,deguchi:04,raghu:10,wang:13}. The multiband nature implies that the gap sizes and phases may be different on different sheets. From the experiments mentioned in the introduction and theoretical analysis \cite{nomura:02,nomura:05,wang:13} it was concluded that the (``active") unhybridized 2D $\gamma$ band has the largest gap. This gap is nodeless but has deep minima in [100] and [010] directions. 
However, this  cannot describe the presence of nearly nodal quasiparticles concluded from 
transport \cite{izawa:01} and thermodynamic \cite{deguchi:04,deguchi:04a} measurements. Theoretical analysis 
\cite{nomura:02,nomura:05,wang:13} suggests that the nodes appear on the  much smaller gaps of the (``passive") hybridized 2D
$\alpha,\beta$ bands and are shifted by $\pi/4$ with respect to the minima on the active bands. From these theoretical and experimental investigations the multiband nodal chiral triplet gap function of \SR~ was established \cite{nomura:02,nomura:05,maeno:12} as $(n=\alpha,\beta,\gamma)$
\bea
\bd_n(\bk)=\Delta^n_0(T)f_n(\bk)\hbz=\Delta_n(\bk)\hbz .
\label{eq:gapdef1}
\eea
The form factors $f_n(\bk)$ contain high Fourier components because of  the sharp minima in $\Delta_n(\bk)$. Here we restrict to the lowest two Fourier components in the expansion of form factors which are already sufficient to fix the qualitative extremal and nodal structure of the three gaps. We use a model where the gap functions on $\alpha,\beta$ bands are degenerate in the limit of vanishing hybridization $V_\bk\rightarrow 0$, this means they will also be degenerate in the orbital basis $a,b$. Explicitly, written separately for the active and passive gaps we have
%\begin{widetext}
\bea
&&
\Delta_{c}(\bk)=\Delta_0
\bigl[\sin k_x(1+A\cos k_y)
%\non\\&&
+i\sin k_y(1+A\cos k_x)\bigr],
\nonumber\\
&&
\Delta_{a,b}(\bk) \equiv \Delta(\bk)
=\Delta'_0
{\Big [}\sin(k_x+k_y)[1+A'\cos (k_x-k_y)]
\non\\
&&
\hspace{3cm}
+i\sin(k_x-k_y)[1+A'\cos(k_x+k_y)]
{\Big ]}.
\non\\
\label{eq:gapdef2}
\eea
%\end{widetext}
%
Both unhybridized $a,b$ and $c$ bands then have the same type of modified nodal chiral p-wave gap function \cite{nomura:02,nomura:05}. The different Fermi surface radii  and the effect of the hybridization will lead to a splitting of gaps of $\alpha,\beta$ bands on the respective Fermi surface sheets.
For $\Delta_0=\Delta'_0$,  $A=A'=0$ and going to the continuum representations one obtains the original chiral p-wave gap $\Delta_n(\bk)=\Delta_0(k_x+ik_y)$ proposed by Rice and Sigrist \cite{rice:95} which is the same on all three bands and has no nodes on the Fermi surface of Fig.~\ref{fig:Fig1}b (the $\pi/4$ rotation of coordinates in the a,b case is implied here). The chiral nature of the gap in Eq.~(\ref{eq:gapdef2}) is compatible with the time-reversal symmetry breaking observed in $\mu SR$ experiments \cite{luke:98}. 

The above model has four parameters, gap amplitudes $\Delta_0$, $\Delta'_0$ and higher harmonic contents $A, A'$.
They may be determined in such a way that we obtain the basic extremal and nodal structure of gap functions $\Delta_n(\theta)$ in Fig.~\ref{fig:Fig2}a. Note that the maxima and minima of active and passive bands are shifted by an angle $\theta_0=\frac{\pi}{4}$ as a consequence of the corresponding \bk- space coordinate rotation in $\Delta_{a,b}(\bk)$.
The parameters of the above model that reproduce the microscopic gap calculation in Refs. \onlinecite{nomura:02,nomura:05} reasonably well are given in the caption of Fig.\ref{fig:Fig2}.

\section{Calculation of Green's functions}
\label{sec:Green}

For the calculation of QPI spectrum we need the Green's function in the superconducting state. The impurity scattering will be treated in Born approximation. This is sufficient if we are not interested in the resonance phenomena associated with strong scattering \cite{akbari:13,akbari:13a}. In the calculation we include all three bands and their active and passive gap functions because they may dominate QPI features for different ranges of the bias voltage V (or frequency $\omega$). For the decoupled nonhybridized single band the expression of the QPI spectrum is known (e.g. Ref. \onlinecite{akbari:13}) and will be added in the end. Here we treat the more involved hybridized subsystem of passive a,b orbitals which will dominate QPI contributions at low frequencies.

Their  projected mean field BCS Hamiltonian is written in $8\times 8$ matrix form in terms of the $8$ - component Nambu spinors
$\Psi_\bk^\dagger = (\psi^\dg_{\bk},\psi_{-\bk})$ with $\psi_{\bk}^\dagger = (c^\dg_{\bk a\ua},c^\dg_{\bk b\ua},
c^\dg_{\bk a\da},c^\dg_{\bk b\da})$ where $a,b$ denote the $xz,yz$ orbitals, respectively. We then have $(n=a,b)$
\bea
H_{SC}&=&\sum_{\bk n\si}
(\varepsilon_{\bk n}-\mu)c_{\bk n\si}^\dagger c_{\bk n\si}
\nonumber\\
&&+
\frac{1}{2}\sum_{\bk n\si\si'}(\Delta_{\bk n}^{\si\si'}c_{-\bk n\si}^\dagger c_{\bk n\si'}^\dagger +H.c.).
\label{eq:HBCS}
\eea
where the gap function $\Delta_{n\bk}=\bd_n(\bk)\cdot\bsig(i\sigma_y)$ is given by Eqs.~(\ref{eq:gapdef1},\ref{eq:gapdef2}) which is of the unitary type with $\bd_n\times\bd_n^* =0$ and $\bd_n(\bk)=\Delta_{n\bk}\hbz$. Here $\bsig =(\sigma_x,\sigma_y,\sigma_z)$ are the Pauli matrices in spin space; we also define the unit as $\sigma_0=I$.

The bare $8\times 8$ Green's function (2 Nambu, 2 orbital, 2 spin degrees of freedom) in the superconducting state is given by
\begin{widetext}
\bea
\hG^{-1}({\bk,\om})=
\left[
 \begin{array}{cccc}
(\om- \epsilon_{a\bk })\sigma_0 & -V_\bk\sigma_0 & -\Delta_{a\bk}\sigma_x & 0\\
-V_\bk\sigma_0 & (\om- \epsilon_{b\bk})\sigma_0  & 0 & -\Delta_{b\bk}\sigma_x\\
-\Delta_{a\bk}^\dagger\sigma_x  & 0 &  (\om +\epsilon_{a\bk})\sigma_0 &V_\bk\sigma_0\\
0 & -\Delta_{b\bk}^\dagger\sigma_x  & V_\bk\sigma_0 &  (\om+ \epsilon_{b\bk})\sigma_0
\end{array}
\right].
\label{eq:Greensinv}
\eea
\end{widetext}
This matrix may be inverted and written in terms of $4 \times 4$ blocks  as
\bea
\hG({\bf  k,\om})=
\left[
 \begin{array}{cc}
 G(\bk,\om)& F(\bk,\om) \\
 F(\bk,\om)^\dagger& -G(-\bk,-\om) 
\end{array}
\right].
\label{eq:Greensmat}
\eea
Here the block index is the Nambu spin $\tau_z$ (2) and each $4\times 4$ block is indexed
by orbital $\kappa_z$ (2) and spin $\sigma_z$ (2) degree of freedom.
The individual blocks may be written as 
\bea
G({\bf  k,\om})&=&
\left[
 \begin{array}{cc}
 G_{aa}(\bk,\om)& G_{ab}(\bk,\om) \\
 G_{ba}(\bk,\om)& G_{bb}(\bk,\om) 
\end{array}
\right]\otimes\sigma_0;% \;\;\;\;
\nonumber\\
F({\bf  k,\om})&=&
\left[
 \begin{array}{cc}
 F_{aa}(\bk,\om)& F_{ab}(\bk,\om) \\
 F_{ba}(\bk,\om)& F_{bb}(\bk,\om) \\
\end{array}
\right]\otimes \sigma_x.
\label{eq:Greensblock}
%\nonumber\\
\eea
We restrict here to the relevant case $\Delta_{a\bk}=\Delta_{b\bk} \equiv \Delta_\bk$ of the \SR~ gap model in
Eqs.~(\ref{eq:gapdef1},\ref{eq:gapdef2}). The general solution will be given in Appendix \ref{sec:appA}. We obtain
for the orbital matrix elements of normal Green's functions:
\begin{widetext}
\bea
G_{aa}(\bk,\om)&=&D(\bk,\om)^{-1}\bigl[(\om-\e_{b\bk})(\om+E_{1\bk})(\om+E_{2\bk})-|\Delta_{\bk}|^2(\om+\e_{a\bk})\bigr]\non\\
G_{bb}(\bk,\om)&=&D(\bk,\om)^{-1}\bigl[(\om-\e_{a\bk})(\om+E_{1\bk})(\om+E_{2\bk})-|\Delta_{\bk}|^2(\om+\e_{b\bk})\bigr]\non\\
G_{ab}(\bk,\om)&=& G_{ba}(\bk,\om)= D(\bk,\om)^{-1}V_{\bk}\bigl[(\om+E_{1\bk})(\om+E_{2\bk})-|\Delta_{\bk}|^2\bigr],
%\non\\
\label{eq:GelementAa}
\eea
and for the anomalous part the result is
\bea
F_{aa}(\bq,\om)&=&D(\bk,\om)^{-1}\Delta_{\bk}\bigl[(\om)^2-E_{b\bk}^2-V_\bk^2\bigr]\non\\
F_{bb}(\bq,\om)&=&D(\bk,\om)^{-1}\Delta_{\bk}\bigl[(\om)^2-E_{a\bk}^2-V_\bk^2\bigr]\non\\
F_{ab}(\bq,\om)&=&F_{ba}(\bq,\om)
=D(\bk,\om)^{-1}\Delta_{\bk}V_\bk(\e_{a\bk}+\e_{b\bk}),
\label{eq:FelementAa}
%\nonumber\\
\eea
%
%Furthermore we have $G_{ba}(\bk,\om)=G_{ab}(\bk,\om)$ and $F_{ba}(\bk,\om)=F_{ab}(\bk,\om)$.
Here the determinant  $D(\bk,\om)$   is given by
\bea
D(\bk,\om)&=&((\om)^2-E_{a\bk}^2)((\om)^2-E_{b\bk}^2)-2V_\bk^2\bigl[
(\om)^2+\bigl(\e_{a\bk}\e_{b\bk}-|\Delta_{\bk}|^2\bigr)\bigr]+V_\bk^4,
\label{eq:GdetAa}
\eea\\
\end{widetext}
where $E_{a\bk}, E_{b\bk}$ are unhybridized quasiparticle energies given by $E_{n\bk}^2=\e_{n\bk}^2+|\Delta_{n\bk}|^2$. The determinant may also be factorized (Eq.~(\ref{eq:Gdetb})) by using the hybridized quasiparticle energies given by 
$\Omega_{1,2\bk}^2=E_{1,2\bk}^2+|\Delta_\bk|^2$ in the case of equal gaps.\\

In the normal state ($\Delta_\bk\equiv 0$) the anomalous Green's function vanishes, i.e., $F_{\alpha\beta}(\bk,\om)=0$ while  the normal Green's function matrix simplifies to
\bea
G_{aa}(\bk,\om)&=&\frac{(\om-\e_{b\bk})}{(\om-E_{1\bk})(\om-E_{2\bk})},
\non\\
G_{bb}(\bk,\om)&=&\frac{(\om-\e_{a\bk})}{(\om-E_{1\bk})(\om-E_{2\bk})},
\non\\
G_{ab}(\bk,\om)&=&G_{ba}(\bk,\om)=\frac{V_\bk}{(\om-E_{1\bk})(\om-E_{2\bk})}.
\label{eq:GelementAb}
\nonumber\\
\eea
Finally when the hybridization vanishes ($V_\bk=0$) then  $G_{nm}(\bk,\om)=\delta_{nm}(\om-\e_{n\bk})^{-1}$ with $n=a,b$ is the usual  normal state unhybridized Green's function matrix. This is equivalent to the c band where  $G_c(\bk,\om)=(\om-\e_{c\bk})^{-1}$.

%%%%%%%%%%%%%%%%%%%%%% figure %%%%%%%%%%%%%%%%%%%%%%%%%%%%%%%%%%%%%%%%%%%%%
\begin{figure}%[t]
\includegraphics[width=0.99\linewidth]{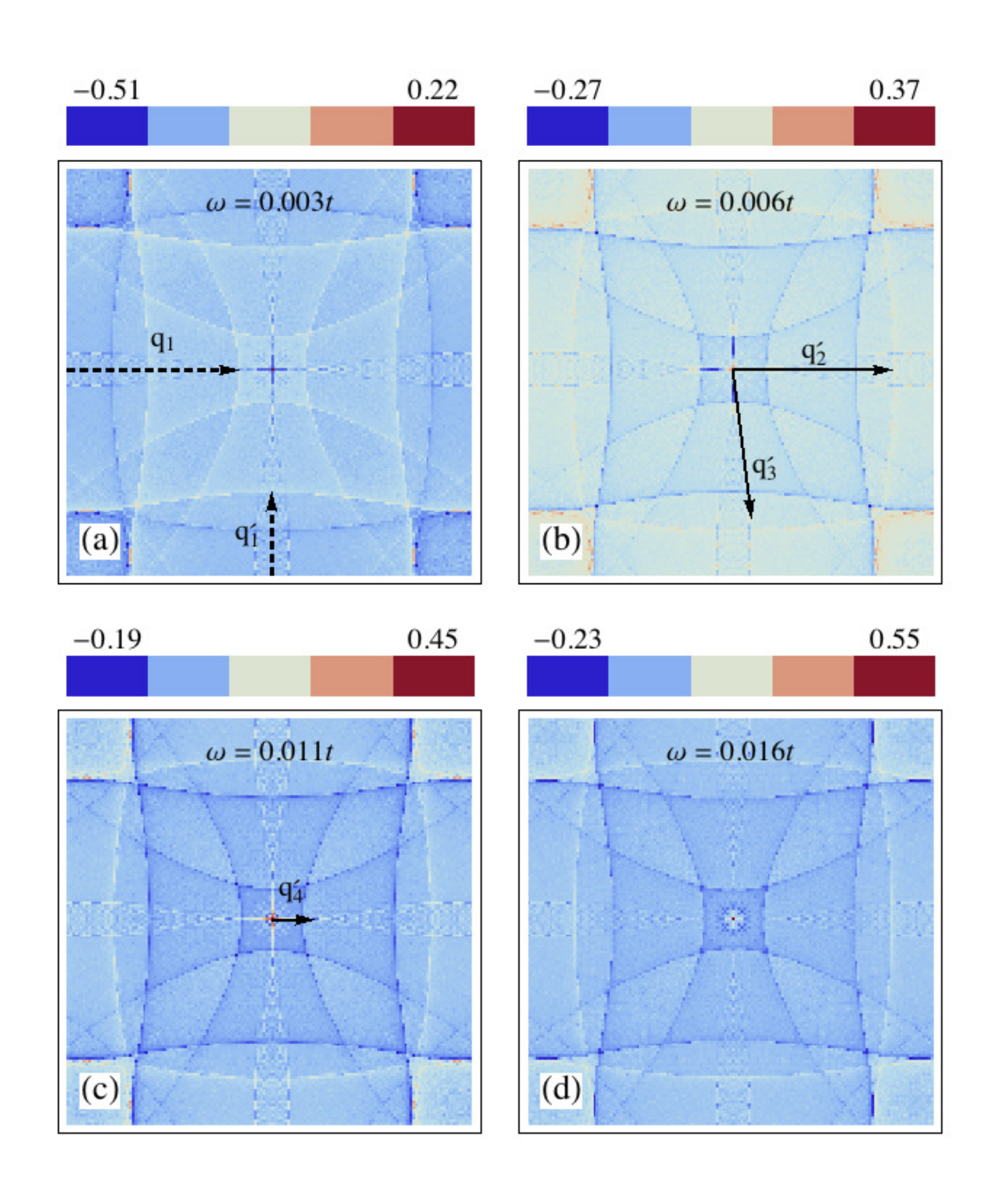}
\caption{QPI spectrum for the normal state. Characteristic QPI wave vectors $\bq_i$ and $\bq'_i$ associated with $\gamma$ and $\alpha,\beta$ bands are indicated (c.f. Fig.~\ref{fig:Fig1}b).
}
\label{fig:Fig4}
\end{figure}
%%%%%%%%%%%%%%%%%%%%%%%%%%%%% figure %%%%%%%%%%%%%%%%%%%%%%%%%%%%%%%%%%%%%%%

\section{Impurity scattering}
\label{sec:impurity}

We describe the effect of normal impurity scattering within the hybridizing a,b subspace. For the single
c orbital results are completely equivalent without involving the trace over orbital subspace. The elastic
scattering potential is given by
\be
\hU(\bq)=[U_c(\bq)\tau_3\sigma_0+U_m(\bq)\tau_0\sigma_z]\ka_0 =\hU_c+\hU_m,
\label{eq:impscatt}
\ee
where we assumed that only intraband scattering $(\sim\ka_0)$ is present.
Here $\sigma,\tau,\kappa$ denote Pauli matrices in spin, Nambu and orbital (a,b) space, respectively.
In Born approximation the full Green's function including the scattering effect is given by $(\bk'=\bk-\bq)$
\bea
\hG_s(\bk,\bk'\om)=\hG(\bk)\delta_{\bk\bk'}+\hG(\bk,\om)\hU(\bq)\hG(\bk',\om).
\non\\
\label{eq:Gfunc}
\eea
The single particle density of states by the scattering is then obtained as (per spin)
\bea
N_s(\bq,\om)
&=&-\frac{1}{\pi}\frac{1}{2N}
{\rm Im}\sum_\bk
{\Big[}
{\rm  tr}_{\si\tau\ka}\hG_t(\bk,\bk'\om)
{\Big ]}
\non\\
&=&N(\om)+\delta N(\bq,\om),
\label{eq:DOS}
\eea
where $N(\om)=(1/N)\sum_{\bk\ka}\delta(\omega-E_{\bk\ka})$ is the background DOS of hybridized bands and
$\delta N(\bq,\om)$ the modification of the local DOS due to impurity scattering. It may be written in terms of
the QPI function $\tL(\bq,\om)$ (for a,b orbitals)  as 
\bea
\delta N(\bq,\om)&=&-\frac{1}{\pi}
{\rm Im}
 \tL_0(\bq,\om)\non\\
\tL_0(\bq,\om)&=&\frac{1}{2N}\sum_\bk {\rm tr}_{\si\tau\ka}\hG_\bk\hU\hG_{\bk-\bq}.
\label{eq:QPI}
\eea
%
%%%%%%%%%%%%%%%%%%%%%% figure %%%%%%%%%%%%%%%%%%%%%%%%%%%%%%%%%%%%%%%%%%%%%
\begin{figure}%[t]
\includegraphics[width=0.99\linewidth]{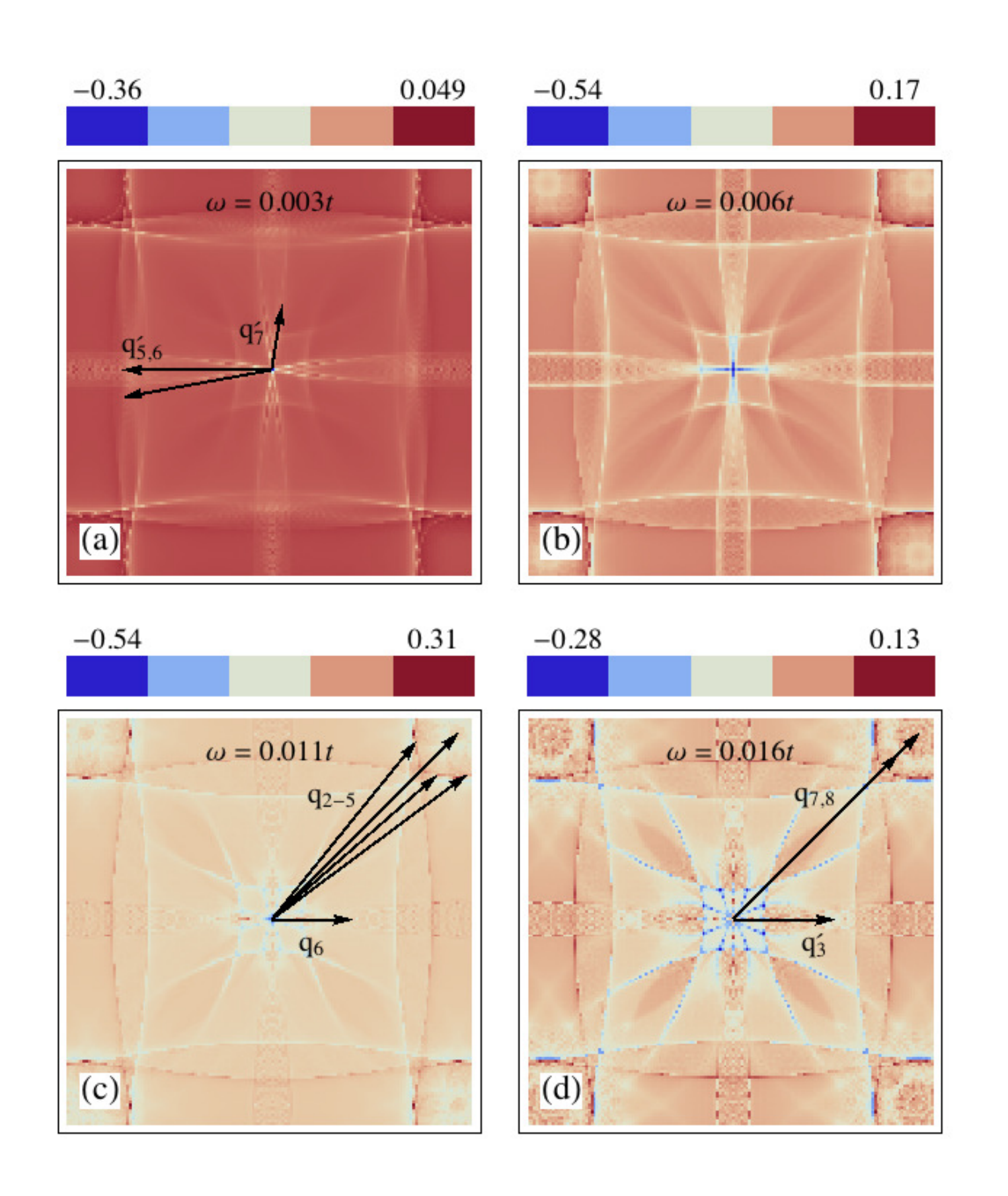}
\caption{QPI spectrum for the superconducting state. Characteristic QPI wave vectors $\bq_i$ and $\bq'_i$ associated with $\gamma$ and $\alpha,\beta$ bands are indicated (c.f. Fig.~\ref{fig:Fig3}).
}
\label{fig:Fig5}
\end{figure}
%%%%%%%%%%%%%%%%%%%%%%%%%%%%% figure %%%%%%%%%%%%%%%%%%%%%%%%%%%%%%%%%%%%%%%

\section{The quasiparticle interference spectrum}
\label{sec:QPI}

The  QPI function in a,b orbital subspace for nonmagnetic scattering ($U_m=0$) in the charge channel
in Born approximation is given by $\tL_0(\bq,\om)=U_c\Lambda'_0(\bq,\om)$ with
\bea
\Lambda'_0(\bq,\om)&=&\frac{1}{2N}\sum_\bk tr_{\si\tau\al}\hG_\bk \tau_3\sigma_0\alpha_0\hG_{\bk-\bq}.
\label{eq:QPItrace}
\eea
The calculation of $\Lambda'_0(\bq,\om)$ may now proceed numerically as is usually done. However here we use the fully analytical closed solution for the QPI spectrum because it gives considerably more insight. In particular the relation to special cases of the model becomes clearer.
For that purpose we perform the traces and use the explicit analytical form of the orbital matrix elements of normal and anomalous Green's functions in Eqs.~(\ref{eq:GelementAa},\ref{eq:FelementAa}). This leads to the QPI function per spin in a,b orbital subspace given by (n,m=a,b):
\bea
\Lambda'_0(\bq,\om)&=&\frac{1}{N}\sum_{\bk,nm}\bigl[
G_{nm}(\bk)G_{nm}(\bk-\bq)
\non\\
&&
\hspace{1.3cm}
-F_{nm}(\bk)F_{nm}(\bk-\bq)^*\bigr].
\label{eq:QPIab}
\eea
To this the contribution of unhybridized c- orbital has to be added which is explicitly given by
\bea
&&\Lambda_0(\bq,\om)=
\non\\
&&\hspace{0.35cm}
\frac{1}{N}\sum_\bk\frac
{(\om+\e_{c\bk})(\om+\e_{c\bk-\bq})-\Delta_{c\bk}\Delta^*_{c\bk-\bq}}
{(\om)^2-(\e_{c\bk}^2+|\Delta_{c\bk}|^2)}.
\label{eq:QPIc}
\eea
The total Born QPI spectrum  $\Lambda^t_0(\bq,\om)=\Lambda_0(\bq,\om)+\Lambda'_0(\bq,\om)$ of active and passive bands, respectively  is then obtained as a closed solution from Eqs.~(\ref{eq:QPIab},\ref{eq:QPIc}) and  Eqs.~(\ref{eq:GelementAa},\ref{eq:FelementAa},\ref{eq:GdetAa}) for the individual matrix elements  $G_{nm}(\bk), F_{nm}(\bk)$. Here we made the simplifying assumption that tunneling matrix elements of a,b and c orbitals are equal.
It is useful to consider the result first for the normal state ($\Delta_\bk=\Delta_{c\bk}=0$) , in this case, using  Eq.~(\ref{eq:GelementAb}) it simplifies to 
\begin{widetext}
\bea
\Lambda^t_0(\bq,\om)&=&\frac{1}{N}\sum_{\bk,n=a,b}\Bigl[\frac
{(\om-\e_{n\bk})(\om-\e_{n\bk-\bq}) +V_\bk V_{\bk-\bq}}
{(\om-E_{1\bk})(\om-E_{2\bk})(\om-E_{1\bk-\bq})(\om-E_{2\bk-\bq})}\Bigr]
+ \frac{1}{N}\sum_{\bk}\Bigl[\frac
{1}{(\om-\e_{c\bk})(\om-\e_{c\bk-\bq})}\Bigr].
\nonumber\\
\label{eq:QPInormal}
\eea
\end{widetext}
It further reduces to $\Lambda^t_0(\bq,\om)=(1/N)\sum_{n\bk}(\om-\e_{n\bk})^{-1}(\om-\e_{n\bk-\bq})^{-1}$ with $n=a,b,c$ for unhybridized bands ($V_\bk=0$). The QPI spectrum in Eq.~(\ref{eq:QPInormal}) is only determined by the dispersion of the three  bands and will map the prominent wave vectors of their corresponding surfaces of constant energy $\omega$.\\

\section{Discussion of numerical results for the three-band chiral gap model}
\label{sec:numerical}

The band structure and associated Fermi surface model for \SR~ is shown in Fig.~\ref{fig:Fig1} consisting of the hybridized $\alpha ,\beta$ and one unhybrized $\gamma$ band. Typical wave vectors  $\bq_i,\bq'_i$  characterising the FS sheet dimensions are indicated (b). These should appear prominently in the normal state QPI functions. We note, however, that the full QPI landscape in $(q_x,q_y)$- space  may not be completely characterised by such characteristic wave vectors and they may not always be unambiguously identified.\\ 

The simplified gap model of Eq.~(\ref{eq:gapdef2}) on this Fermi surface is shown in Fig.~\ref{fig:Fig2}a. It reproduces the overall extremal and nodal behaviour obtained by a fully microscopic model in Refs.~\onlinecite{nomura:02,nomura:05}, in particular the shifted minima or nodes  of the gap functions on active ($\gamma$) and passive ($\alpha,\beta$) bands. Since the model of Eq.~(\ref{eq:gapdef2}) includes only two Fourier components for each band there are, however, quantitive differences to the full calculation in  Refs.~\onlinecite{nomura:02,nomura:05}. This has little influence on the overall appearance  of the QPI spectra. The associated quasiparticle DOS for this gap model is shown in  Fig.~\ref{fig:Fig2}b. Note that the $\gamma$-band DOS and therefore the total DOS are slightly asymmetric. This is due to
the behaviour of normal state DOS around the Fermi level. It is determined by the asymmetric behaviour of the $\gamma$ band dispersion around the X -point (Fig.~\ref{fig:Fig1}a).\\

The features of  QPI functions in the SC state are determined by the shape of surfaces of constant quasiparticle energy surfaces given by $\Omega_{n\bk}=\omega$  $(n=1-3)$. They are shown in Fig.~\ref{fig:Fig3}. For small $\omega \ll \Delta'_0$ first the double arc-shaped sheets around the [110]-type nodal directions of the $\alpha,\beta$ bands appear (a). 
For $\omega > \Delta'_0/2$ one basically obtains the (doubled) constant energy surface sheets of the normal state (b). When the frequency $\omega$ increases above the minimum of the $\gamma$- band,  lens-shaped $\gamma$- sheets around the [100]-type extremal directions appear which are rotated by $\pi/4$ with respect to low energy $\alpha-\beta$ arc-shaped sheets (c). Finally when $\omega$ is above the maximum gap in Fig.~\ref{fig:Fig2} doubled normal constant energy surface sheets split by the gap appear also for the $\gamma$ band.
 The prominent connecting wave vectors of those sheets, if they appear in the QPI spectrum, should give information on Fermi surface structure and in the superconducting state a direct evidence for the nodal structure of the gap function. Several candidate wave vectors are indicated in Fig.~\ref{fig:Fig3}.
For clarity we denote by $\bq_i$ and $\bq'_i$ $(i=1,2,3 ..)$ wave vectors connecting points on equal energy surfaces of $\gamma$ and $\alpha,\beta$ bands, respectively. We do not distinguish between wave vectors related by fourfold symmetry.\\

First we discuss the normal state in  Fig.~\ref{fig:Fig4} where we show the QPI spectrum for four increasing energy values (c.f. Fig.~\ref{fig:Fig2}a and Fig.~\ref{fig:Fig3}). At the wave vectors $\bq_1, \bq'_1,\bq'_2$ associated with the main across-Fermi surface scattering processes clearly line structures are seen in the QPI spectrum at all energies $\omega$. Note that $\bq_1, \bq'_1$ are folded back into the first BZ. To compare with the vectors in  Fig.~\ref{fig:Fig1}b one has to add zone boundary vectors $(\pi,0)$ and $(0,\pi)$, respectively. There are also weaker lines emanating from zone center and forming a split cross, in particular visible in (Fig.~\ref{fig:Fig4}c). They can be associated parallel scattering along $\alpha,\beta$-sheets including hybridization-induced interband-scattering between $\alpha,\beta$ bands. It will appear according to Eq.~(\ref{eq:GelementAb}) although there is no interband scattering potential.
The split crosses are obtained by tracking wave vectors of the type $\bq'_3$ (and the one reflected at the symmetry plane)  in Fig.~\ref{fig:Fig1}b from zero to the zone boundary. Furthermore the small wave vector axis-aligned cross features in Fig.~\ref{fig:Fig4} are due to scattering parallel to $\alpha\beta$ surfaces with wave vector $\bq'_4$.  Therefore all major features observed in normal state QPI of Fig.~\ref{fig:Fig4} can be reasonably understood from the hybridized three-band Fermi surface structure.\\

Now we turn to the QPI in the chiral p-wave superconducting state described by Eq.~(\ref{eq:gapdef2}).
By tuning the bias voltage or frequency it is clear that the most significant information on the gap function may be obtained in situations like Fig.~\ref{fig:Fig3}a,c where the small arc and lens shaped sheets first appear around the nodal or extremal directions, respectively.
These small sheets have points of high curvature and may show up as distinguished features in the QPI. 

In Fig.~\ref{fig:Fig5}a,b the QPI signature of the small gap  on the  $\alpha, \beta$ bands at $\bq'_5-\bq'_7$ (Fig.~\ref{fig:Fig3}a) are apparently rather weak. This is due to the smallness of the gap $\Delta'_0/t=10^{-2}$. However clearly the intensities at $\bq'_{5,6}$ as compared to neighboring wave vectors is enhanced with respect to the normal state. The situation here is quite different from the heavy fermion system \CC~\cite{akbari:11} where the gap is only about one order of magnitude less than the effective hopping. Then the QPI in the SC state shows up more clearly.

This situation changes when the energy is raised to the region of the large gap on the active $\gamma$ sheet. The scattering vectors connecting the lens- shaped $\gamma$ surface sheets in Fig.~\ref{fig:Fig3}c at wave vectors $\bq_2 -\bq_5$ clearly turn up as separate features in the QPI of Fig.~\ref{fig:Fig5}c. They partly survive to even higher energy in Fig.~\ref{fig:Fig5}d when the constant energy surfaces of the $\gamma$ band are already reconnected again (Fig.~\ref{fig:Fig3}d). The observation of this wave vector quadruplet  $\bq_2 -\bq_5$ above some threshold energy $\omega_0$ would be a clear indication of the active gap having a minimum of the size $\Delta_{min}\simeq \omega_0$ in the [100] -type directions.
In the low momentum region of Fig.~\ref{fig:Fig5}c it is also possible to identify the intra-lens scattering vector $\bq_6$ of  Fig.~\ref{fig:Fig3}c . Furthermore the vector $\bq'_3$  in Fig.~\ref{fig:Fig5}d is apparently related to the axis parallel scattering in Fig.~\ref{fig:Fig3}d made possible by the doubling of $\alpha,\beta$ sheets in the superconducting state.

\section{Conclusion and outlook}
\label{sec:conclusion}

In this work we investigated the QPI spectrum of multi-band chiral p-wave superconductor \SR. Our working model which is a simplified verison of the microscopic three-band model studied first in Refs.~\cite{nomura:02,nomura:05} and experimentally vindicated in Refs.\onlinecite{deguchi:04a,maeno:12}. It consists of  a non-hybridized $xy$ type active $\gamma$ band with a Fermi surface that supports the main chiral gap function. The latter has deep minima along [100] type directions.  A secondary near nodal gap is supported by the hybridizing 
$xz,yz$ type $\alpha, \beta$ bands with a near nodal structure that is rotated by $\pi/4$ with respect to the minima of the large $\gamma$ band gap. 
In the normal state the basic across -Fermi surface scattering appears as clear line features in the QPI spectrum originating from all three bands. The QPI changes due the $\alpha,\beta$ gap opening are quite subtle due to the smallness of the gap and they are caused by low momentum scattering between the arc-shaped $\alpha-\beta$ surfaces in Fig.~\ref{fig:Fig3}a.\\

A more clear signature in QPI is left by the dominant chiral p-wave gap on the $\gamma$ surface. The scattering due to lens- type constant energy surfaces  (Fig.~\ref{fig:Fig3}c) around the gap minima positions along [100] direction leads to a quadruplet of wave vectors that can be identified in QPI spectrum. Its observation would support the existence of the minimum in the main gap.
Further less prominent wave vectors may also be identified in the QPI structure.

In general the QPI analysis should be focused in those voltage regions where equal energy surfaces have the shape as in Fig.~\ref{fig:Fig3}a,c. Outside these regions the equal energy surfaces are, aside from the doubling quite similar to the normal state (Fig.~\ref{fig:Fig3}a,c) and then little change may be expected. It would be most interesting to see whether QPI can confirm the relative $\pi/4$ rotation of (near) nodal positions on $\alpha-\beta$ and extremal positions $\gamma$ band from the characteristic wave vectors in Fig.~\ref{fig:Fig3}a,c and Fig.~\ref{fig:Fig5}a,c.
Our results suggest that it is worthwhile to investigate \SR~ by QPI method to learn more about its electronic structure and in particular the chiral p-wave gap function.

\appendix
\section{}
\label{sec:appA}

In this Appendix we discuss the most general case of the QPI when the gap functions for hybridizing orbitals $\Delta_{a\bk},\Delta_{b\bk}$ may be unequal. Generally this implies a breaking of fourfold symmetry of QPI spectra in the tetragonal plane when  the gap amplitudes $\Delta'_{n0}$ for $n=a,b$ are different. There is no evidence of spontaneous fourfold symmetry breaking in the superconduting phase of \SR~ from field-angle dependent specific heat analysis \cite{deguchi:04}. Nevertheless we include this case here because it may be useful for other multiband superconductors. We obtain
\begin{widetext}
\bea
G_{aa}(\bk,\om)&=&D(\bk,\om)^{-1}\bigl[(\om-\e_{b\bk})(\om+E_{1\bk})(\om+E_{2\bk})-|\Delta_{b\bk}|^2(\om+\e_{a\bk})\bigr]\non\\
G_{bb}(\bk,\om)&=&D(\bk,\om)^{-1}\bigl[(\om-\e_{a\bk})(\om+E_{1\bk})(\om+E_{2\bk})-|\Delta_{a\bk}|^2(\om+\e_{b\bk})\bigr]\non\\
G_{ab}(\bk,\om)&=&D(\bk,\om)^{-1}V_{\bk}\bigl[(\om+E_{1\bk})(\om+E_{2\bk})-\Delta_{a\bk}\Delta_{b\bk}^*\bigr]\non\\
G_{ba}(\bk,\om)&=&D(\bk,\om)^{-1}V_{\bk}\bigl[(\om+E_{1\bk})(\om+E_{2\bk})-\Delta_{a\bk}^*\Delta_{b\bk}\bigr],
\label{eq:Gelement}
\eea
where the determinant is given by
\bea
D(\bk,\om)&=&((\om)^2-E_{1\bk}^2)((\om)^2-E_{2\bk}^2)
-|\Delta_{a\bk}|^2((\om)^2-\e_{b\bk}^2)-|\Delta_{b\bk}|^2((\om)^2-\e_{a\bk}^2)\non\\
&&+(\Delta_{a\bk}^*\Delta_{b\bk}-\Delta_{a\bk}\Delta_{b\bk}^*)V_{\bk}^2+|\Delta_{a\bk}|^2|\Delta_{b\bk}|^2,
\label{eq:Gdet}
\eea
or equivalently it can be expressed as
\bea
D(\bk,\om)&=&((\om)^2-E_{a\bk}^2)((\om)^2-E_{b\bk}^2)-2V_{\bk}^2\bigl[
(\om)^2+\bigl(\e_{a\bk}\e_{b\bk}-\frac{1}{2}(\Delta_{a\bk}^*\Delta_{b\bk}+\Delta_{a\bk}\Delta_{b\bk}^*)\bigl)\bigr]+V_{\bk}^4.
\label{eq:Gdeta}
\eea
Likewise the orbital matrix elements of the anomalous Green's function are obtained as 
\bea
F_{aa}(\bq,\om)&=&D(\bk,\om)^{-1}\bigl[\Delta_{a\bk}((\om)^2-E^2_{b\bk})-\Delta_{b\bk}V_\bk^2\bigr]\non\\
F_{bb}(\bq,\om)&=&D(\bk,\om)^{-1}\bigl[\Delta_{b\bk}((\om)^2-E^2_{a\bk})-\Delta_{a\bk}V_\bk^2\bigr]\non\\
F_{ab}(\bq,\om)&=&D(\bk,\om)^{-1}V_\bk\bigl[-\Delta_{a\bk}(\om-\e_{b\bk})+\Delta_{b\bk}(\om+\e_{a\bk})\bigr]\non\\
F_{ba}(\bq,\om)&=&D(\bk,\om)^{-1}V_\bk\bigl[\Delta_{a\bk}(\om+\e_{b\bk})-\Delta_{b\bk}(\om-\e_{a\bk})\bigr].
\label{eq:Felement}
\eea
The determinant may also be written by using the true quasiparticle energies $\Omega_{1,2}(\bk)$ in the hybridized and superconducting case according to
\bea
D(\bk,\om)&=&((\om)^2-\Omega_{1\bk}^2)((\om)^2-\Omega_{2\bk }^2),
\label{eq:Gdetb}
\eea
with
\bea
\Omega_{1\bk}^2&=&\frac{1}{2}(E_{a\bk}^2+E_{b\bk}^2)+V_\bk^2-
\frac{1}{2}\bigl[(E_{a\bk}^2-E_{b\bk}^2)^2+4V_\bk^2\bigl((\e_{a\bk}+\e_{b\bk})^2 
+|\Delta_{a\bk}-\Delta_{b\bk}|^2\bigr)\bigl]^\frac{1}{2}\non\\
\Omega_{2\bk}^2&=&\frac{1}{2}(E_{a\bk}^2+E_{b\bk}^2)+V_\bk^2+
\frac{1}{2}\bigl(E_{a\bk}^2-E_{b\bk}^2)^2+4V_\bk^2\bigl((\e_{a\bk}+\e_{b\bk})^2 
+|\Delta_{a\bk}-\Delta_{b\bk}|^2\bigr)\bigl]^\frac{1}{2}.
\eea
For the hybridized case
with equal gaps ($\Delta_{a\bk}=\Delta_{b\bk}=\Delta_\bk$) the above equation simplifies to
$\Omega_{1,2\bk}^2=E_{1,2\bk}^2+|\Delta_\bk|^2$.
\end{widetext}

%\newpage
%%%%%%%%%%%%%%%%%%%%%%%%%%%%%      References        %%%%%%%%%%%%%%%%%

\bibliography{References}

\end{document}